# The de Haas-van Alphen quantum oscillations in the kagome metal RbTi$_3$Bi$_5$

Zixian Dong[1], Lei Shi[1], Bing Wang[1], Mengwu Huo[1], Xing Huang[1], Chaoxin Huang[1], Peiyue Ma[1], Yunwei Zhang[1,*], Bing Shen[1,§], and Meng Wang[1,†]

[1]Center for Neutron Science and Technology, Guangdong Provincial Key Laboratory of Magnetoelectric Physics and Devices, School of Physics, Sun Yat-Sen University, Guangzhou, Guangdong 510275, China

\* _zhangyunw@mail.sysu.edu.cn_

§ _shenbing@mail.sysu.edu.cn_

† _wangmeng5@mail.sysu.edu.cn_

ABSTRACT

Kagome system usually attracts great interest in condensed matter physics due to its unique structure hosting various exotic states such as superconductivity (SC), charge density wave (CDW), and nontrivial topological states. Topological semimetal RbTi$_3$Bi$_5$ consisting of the kagome layer of Ti shares a similar crystal structure to topological correlated materials $A$V$_3$Sb$_5$ ($A$ = K, Rb, Cs) but with the absence of CDW and SC. Systematic de Haas-van Alphen (dHvA) oscillation measurements are performed on the single crystals of RbTi$_3$Bi$_5$ to pursue nontrivial topological physics and exotic states. Combining with theoretical calculations, detailed Fermi surface topology and band structure are investigated. A two-dimensional (2D) Fermi pocket β is revealed with a light-effective mass in consistent with the semimetal predictions. Landau Fan of RbTi$_3$Bi$_5$ reveals a zero Berry phase for the β oscillation in contrast to that of CsTi$_3$Bi$_5$. These results suggest the kagome RbTi$_3$Bi$_5$ is a good candidate to explore nontrivial topological exotic states and topological correlated physics.

I Introduction

The kagome lattice with corner-sharing triangles naturally has relativistic band crossings at the Brillouin zone (BZ) corners and usually hosts various exotic states [1-3]. Recently two-dimensional kagome superconductors $A$V$_3$Sb$_5$ ($A$ = K, Rb, Cs) were found and attracted significant interest in condensed matter physics. Theoretical and experiment investigations reveal a Z$_2$ nontrivial topological band structure with multiple protected Dirac crossings and van Hove singularities (vHSs) [4-9]. Around 80 to 100 K, the system undergoes a chiral charge density wave transition with a orbital moment [6-10]. Accompanying this transition, prominent anomalous Hall effect and planar Hall exhibit unconventional behaviors [11-13]. Besides this first order transition, nuclear magnetic resonance (NMR) and muon spin relaxation (μsR) measurements reveal other transitions below CDW transition with a sharp increase of the electronic magnetic chiral anisotropy [14-17]. With applying the pressure, two superconducting regions were identified with multiple exotic orders, suggesting a correlation effect in this nontrivial topological system [18-20].

To pursue the origin of these exotic orders and the interplay of correlation effect and nontrivial topology, various experiments were performed revealing a complicated electronic phase diagram [21,22]. The chiral CDW and SC orders are sensitive to the pressure and chemical doping [23-26]. The observed multiple modulations suggest the emergence of various CDW phases and the electronic nematic states. Two domes of SC were observed in the phase diagram and considered to originate from different mechanisms. Up to now, the interplay of SC and CDW is still puzzling and challenging. The analogue materials $A$Ti$_3$Bi$_5$ ($A$ = Rb, Cs) share a similar crystal structure to $A$V$_3$Sb$_5$ ($A$ = K, Rb, Cs) with a kagome layer made of Ti [27,28]. Theoretical calculations and some experimental results yielded a similar nontrivial topological band structure [29-32]. However, the experimental studies suggested the absence of the chiral CDW [33] in $A$Ti$_3$Bi$_5$ ($A$ = Rb, Cs=and revealed different electron phase diagrams compared to $A$V$_3$Sb$_5$ ($A$ = K, Rb, Cs) [33]. Thus $A$Ti$_3$Bi$_5$ ($A$ = Rb, Cs) can provide a new view to explore the mechanism of electronic nematicity and its interplay with the orbital degree of freedom in these kagome compounds [34,35]. The detailed investigations for band structure and Fermi topology are greatly desired for a better understanding of the interplay between electron magnetism and lattice in a topological correlated system.

In this work, we perform systematic magnetization measurements on the single crystals of RbTi$_3$Bi$_5$. Above 7 T, prominent dHvA oscillations are observed and a light effective mass is revealed. In contrast to CsTi$_3$Bi$_5$, only one frequency of oscillation is identified. Combining with theoretical calculations and angular dHvA oscillation measurements, 2D Fermi surface topology is characterized. The Landau fan diagram reveals a zero Berry phase for this oscillation. This is consistent with the theoretical predication of a Z$_2$ nontrivial topological phase. The absence of CDW suggests different correlation effects in this topological kagome system.

II Experiments

Single crystals of RbTi$_3$Bi$_5$ were synthesized using the self-flux method [38]. In an argon glovebox, Rb (liquid, Alfa 99.98%), Ti (powder, Alfa 99.99%), and Bi (powder, Alfa 99.99%) with the molar ration of Rb: Ti: Bi = 1: 2: 6 were put in a crucible, and then the crucible was sealed in a vacuumed quartz tube. The mixture was first heated to $1000°C$ for 48 hours and cooled down to $600°C$ with $20\,\text{min}/°C$. the crystals were separated from the flux by centrifuging the mixture. The crystal structure was investigated by X-ray diffraction. Electrical transport measurements were carried out in a physical property measurement system (Quantum Design). The magnetization measurements were based on a vibrating sample magnetometer. The density functional theory (DFT) calculations were performed by using Vienna *ab* initio package (VASP). The generalized approximation according to the Perdew-Burke-Ernzerhof type was used to describe the exchange-correlation function. The lattice relaxations were performed until the force was less than 0.001 eV/Å. The plane-wave cutoff energy was 600 eV and the *k* mesh was 20×20×12 in the optimization and self-consistent calculations. The Fermi surfaces were generated using a dense *k*-mesh grid of 20×20×20 and visualized using the XCRYSDEN software package. The quantum oscillation frequencies were calculated using the SKEAF code.

III Results and Discussions

RbTi$_3$Bi$_5$ exhibits a layered crystal structure consisting of Ti-Bi slabs intercalated by Rb cations as shown in Fig. 1(a) and Fig. 1(b). The Bi sublattice forms a perfect 2D kagome net. Our measurements are based on the planar single crystals. The (001) diffraction peaks are revealed by the X-ray diffraction measurements indicating the high-quality crystalline for our sample. The temperature-dependent resistivity $\rho(T)$ shown in Fig. 1(c) exhibits a metallic behaviour with the absence of the anomaly for CDW transition. Correspondingly, the temperature-dependent susceptibility $\chi(T)$ (defined as $M/H$) exhibits a paramagnetic behavior, and no phase-transition anomaly is observed during the whole temperature region in Fig. 1(d). These results suggest that the absence of CDW in RbTi$_3$Bi$_5$ is consistent with the former report. Besides, no SC is observed in $\rho(T)$ and $\chi(T)$ as decreasing the temperature down to 2 K. The absence of CDW and SC compared to $A$V$_3$Sb$_5$ motivated us to investigate the electronic structures of RbTi$_3$Bi$_5$. The DFT calculations reveal several nontrivial topological band structures in Fig. 1(e) and Fig. 1(f). The flat band as a typical character of the electronic structure of a kagome lattice, emerges around a binding energy of 0.3 eV. At the K point, a type-II nodal line is observed. The vHS emerging around the M point is close to the Fermi level ($E_F$). These exotic electronic features are very similar to those in $A$V$_3$Sb$_5$ compounds. Whereas, the obvious difference between $A$V$_3$Sb$_5$ and RbTi$_3$Bi$_5$ would motivate us to further check these theoretical results from the experimental point of view to pursue the different correlation effects in the kagome topological system.

Compared to the Shubnikov-de Haas (electronic transport) quantum oscillation measurements, the dHvA (magnetic) quantum oscillation measurements are insensitive to extrinsic surface states and conditions and provide the Fermi topology information from bulk. In our experiments, the planar single crystal of RbTi$_3$Bi$_5$ was mounted in a wedge-shaped quartz holder with various thicknesses to control the sample plane direction. The isothermal magnetizations of RbTi$_3$Bi$_5$ were measured with an applied field (B) parallel to the $c$ axis. At 2 K, the quantum oscillations were observed above 7 T. After subtracting the polynomial background, remarkable periodic oscillations in $\Delta M(H)$ vs $1/\mu_0 H$ curves are more obviously shown in Fig. 2(a). With increasing the temperature these oscillations become invisible gradually and disappear around 20 K eventually. To analyze these oscillations, the fast Fourier transform (FFT) is performed and related spectra are depicted in Fig. 2(b). An obvious resonance peak is identified at 190 T, consistent with previous Shubnikov-de Haas quantum oscillation results [33]. Besides, a weak resonance peak is identified around a smaller frequency of 12 T after careful checking by subtracting various polynomial backgrounds. According to the Lifshitz-Kossevich (L-K) formula, the oscillatory $\Delta M$ can be expressed as [39]

$$\Delta M \propto \sqrt{B} R_T R_D R_S \sin\left[2\pi\left(\frac{F}{B} - \frac{1}{2} + \frac{\phi_B}{2\pi} - \delta\right)\right]$$

where, $R_T = (\alpha m^* T / \mu_0 B) / \sinh(\alpha m^* T / \mu_0 B)$, $R_D = \exp(-\alpha m^* T_D / \mu_0 B)$,

$R_S = \cos(\pi m^* g^*)$, the parameter $\alpha$ is a constant defined as $\alpha = 14.69 T/K$, besides, $m^*$ is the cyclotron effective mass, $T_D$ is the Dingle temperature, $g^*$ is the effective $g$ factor, $\delta$ is the phase shift, and $\delta$ is 0 for a 2D system and $\pm 1/8$ for a three-dimensional (3D) system. The effective mass $m^*$ for the oscillation of 190 T is calculated to be 0.29 $m_e$. The light effective mass indicates a typical semimetallic band dispersion agreeable with our calculations below and close to that in $CsTi_3Bi_5$ [36]. The Dingle temperature $T_D$ at 2 K is calculated to be 7.164 K, and the corresponding mean-free path is yielded to be 514.024 nm. More detailed results are summarized in TABLE 1.

Angular dHvA quantum oscillation measurements were performed at 2 K to investigate the detailed Fermi topology shown in Fig. 3(a). To map a fine structure of the Fermi surface, the magnetic field is tilted from $c$ axic to the the [110] direction by rotating the sample. The angle between the magnetic field and the c axis defined as θ. Thus, $\theta = 0°$ denotes the field along the $c$ direction and $\theta = 90°$ denotes a field along the [110] direction. According to the Onsager rule, the Fermi surface area and frequencies of quantum oscillation in magnetic fields can be described as $F = \left(\frac{\hbar}{2\pi e}\right) A$, where $A$ is Fermi surface area and $F$ is the frequency of quantum oscillation. The FFT spectra of $\Delta M(H)$ at various angles taken at 2 K are plotted in Fig. 3(b). These oscillations exhibit strong angle dependence and become invisible gradually with rotating the applied field from the $c$ axic to $ab$ plane. The oscillation at F = 190 T varnishes for $\theta > 50°$. By using $F_0 / \cos\theta$ ($F_0$ = 190 T), the angular dependent $F$ can be fitted indicating a typical 2D character for this revealed Fermi surface. To further understand this Fermi topology, DFT calculations were performed to investigate more details of the electronic structure. As shown in Fig. 3(c), multiple Fermi pockets are revealed. Four dispersive bands crossing $E_F$ labeled α to δ, forming one circle-like (α band) and one hexagonal-like (β band) electron pocket. The γ band crosses through $E_F$ twice, which forms a hexagonal-like electron pocket around the Brillouin zone (BZ) center and a triangle-like hole pocket around the BZ corner. Additionally, the δ band constitutes a rhombic-like hole pocket around the boundary of BZ. By analyzing these Fermi sheets, the observed oscillation of 190 T is related to the electron δ sheet. This vessel-shaped Fermi sheet is located around M point exhibiting a strong 2D character. As shown in Fig. 3(e), the calculated oscillatory frequencies of the δ sheet are consistent with the former experimental results. Correspondingly, the effective mass for the δ bands from the *3d* orbit of Ti is calculated to be 0.24 $m_e$ which is close to the experimental value. The failure to observe quantum oscillations for the other Fermi pockets could be attributed to the large moving orbits of the extremal cross-section, which may need further higher field measurements.

To obtain the exact Berry phase, it is crucial to correctly assign the Landau index to the $\Delta M(H)$ extremum. Based on our experimental results, the minima of $\Delta M$ are assigned to be the integer indices and the maxima are assigned to the half-integer indices. The related Landau level fan diagram for the frequency $F_\beta$ at 2 K is determined and presented in Fig. 4. By fitting the Landau index N as a function of $1/\mu_0 H$ at various angles, we obtained Berry phase $\phi_B = 2\pi\left(n_0 - \delta + \frac{1}{2}\right)$. Correspondingly, the angular-dependent Berry phase $\phi_B$ is around 0 and exhibits angular independence shown in the inset in Fig. 4. The DFT calculation reveals three types of nontrivial topological band structures for RbTi$_3$Bi$_5$. At the K point, the type-II Dirac nodal line is protected by a mirror symmetry if without considering the spin-orbital coupling (SOC). A negligible gap opens at the type-II nodal line by including the SOC, which is distinct from the pronounced gaps that Dirac point open at the trivial band-crossing points. This kind of nontrivial topological band structure usually can host additional Berry curvature resulting in a π shift in Berry phase for quantum oscillations such as that in CsTi$_3$Bi$_5$. The $\phi_B = 0$ in RbTi$_3$Bi$_5$ may be attributed to the position of Dirac nodal line a little far away from the Fermi level. On the other hand, around M point, the vHS seems insensitive to additional Berry curvature.

Our experimental and theoretical investigations on RbTi$_3$Bi$_5$ reveal similar electronic structures as RbV$_3$Sb$_5$. However, the DFT calculations suggest that the vHSs at the M points are a little far away from $E_F$. The imperfect electronic nesting seems to fail to drive CDW order. Recent research suggested even tuning the vHS by doping in $A$Ti$_3$Bi$_5$, the electronic nesting between the vHSs at the M point alone is also insufficient to drive a CDW order in these kagome metals [37]. The absence of CDW could be attributed to the different correlation effects or structure instability in various kagome systems.

IV CONCLUSION

In summary, we have investigated the electronic structure of single crystals of RbTi$_3$Bi$_5$ by de Haas-van Alphen oscillations. Combined with DFT calculations, four types of Fermi surfaces are revealed. The δ Fermi pocket related to a 190 T quantum oscillation β revealed experimentally exhibits a 2D character. The Landau level fan diagram reveals Berry phase of the β oscillation in RbTi3Bi5 is around zero in contrast to that with a π shift in CsTi3Bi5. The results suggest RbTi3Bi5 is a sensitive candidate material to explore exotic electronic states and topological correlated physics in kagome system.

ACKNOWLEDGEMENTS

This work was supported by the he National Key Research and Development Program of China (Grant Nos. 2023YFA1406500), National Natural Science Foundation of China (grant no. 12174454), the Guangdong Basic and Applied Basic Research Funds (grant no. 2021B1515120015), Guangzhou Basic and Applied Basic Research Funds (grant nos. 2024A04J6417), Guangdong Provincial Key Laboratory of Magnetoelectric Physics and Devices (grant no. 2022B1212010008), the National Natural Science Foundation of China (Grant Nos. U2130101 and 92165204), Natural Science Foundation of Guangdong Province (Grant No. 2022A1515010035).

| | $F$ | $A$ | $k_F$ | $m^*$ | $T_D$ | $l_q$ | $\tau_s$ | $\mu$ |
| | $T$ | $nm^{-2}$ | $nm^{-1}$ | $m_e$ | $K$ | $nm$ | $s$ | $m^2/Vs$ |
|---|---|---|---|---|---|---|---|---|
| $H//C$ | 190.7 | 1.816 | 0.760 | 0.291 | 7.164 | 514.024 | $1.70 \times 10^{-13}$ | 1.026 |

TABLE 1. Physical parameters: The frequency ($F$), cross-section area ($A$), Fermi wave vector ($k_F$), effective mass ($m^*$), Dingle temperature ($T_D$), mean-free path ($l_q$), quantum relaxation time ($\tau_s$), and quantum mobility ($\mu$) characterizing the dHvA oscillation of RbTi$_3$Bi$_5$.

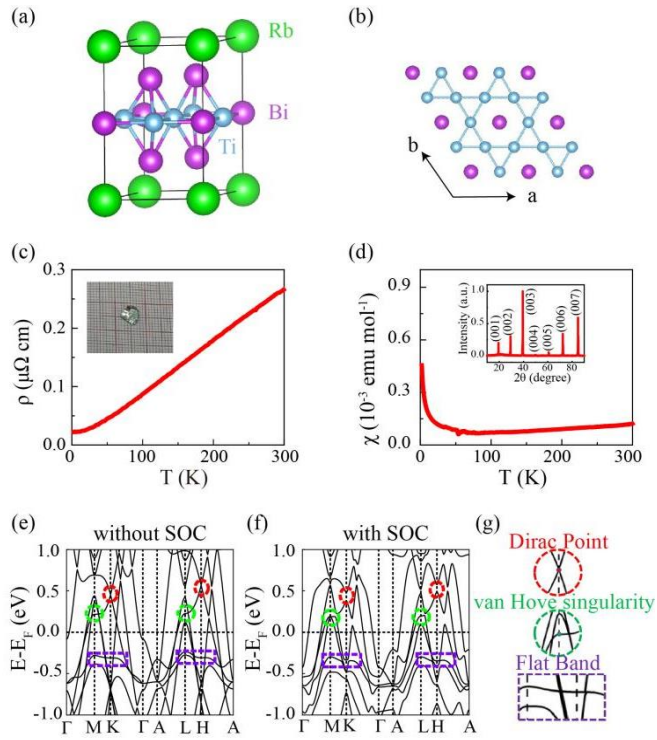

FIG 1. (a) A sketch of the crystal structure of RbTi$_3$Bi$_5$. (b) A two-dimensional kagome plane consisting of Bi and Ti. (c) Temperature dependence of the resistivity and inset is a photo of a typical single crystal. (d)Temperature dependence of magnetic susceptibility with μ$_0$H = 1 T along *c* axis during zero field cooling. The inset shows an X-ray diffraction pattern of the single crystal of RbTi$_3$Bi$_5$. (e) and (f) Band structures of RbTi$_3$Bi$_5$ without SOC (e) and with SOC (f). (g) The band structure of Dirac Point (DP), van Hove singularity (vHS), and flat bands are marked as red, green, and purple circles, respectively.

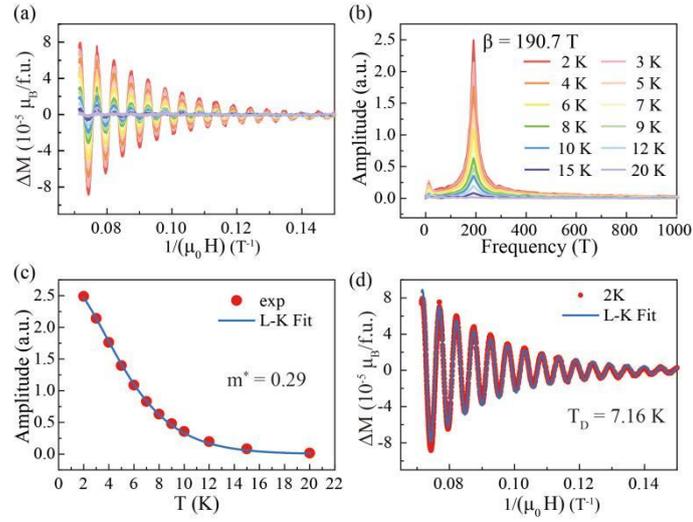

FIG 2. (a) Magnetization after subtracting by a monotonic background ΔM as a function of $1/\mu_0 H$ at various temperatures with a magnetic field along the *c* axis. (b) Fast Fourier Transforms (FFT) spectra of ΔM. The oscillation frequency is donated as *β*. (c) Temperature dependence of the FFT amplitudes of the oscillations in (b). The solid curve represents the best fit to the data using the L-K formula, where the effective mass *m\** can be obtained. (d)The oscillation of the subtracted magnetization at 2 K. The red line is the experimental data and the blue line is the L-K fit. The Dingle temperature $T_D$ is 7.1 K.

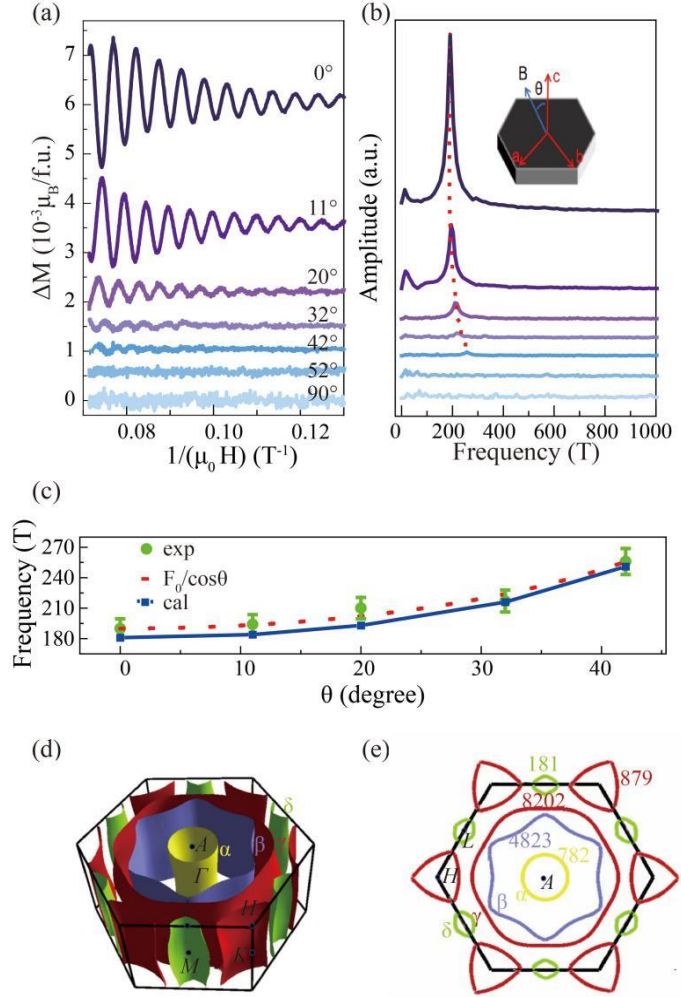

FIG. 3. (a) dHvA oscillation as a function of $1/\mu_0H$ at various angles. (b) FFT spectra of the dHvA oscillation at different angles. The inset picture shows the orientation of the magnetic field. The red dotted line is a guide to the eye. (c) Angle dependence of the FFT frequencies. The green point is the experimental results with a 5 % error. The red dotted line and the blue solid line represent the $F_0/\cos\theta$ fitting line and the values of theoretical calculations at various angles, respectively. (d) 3D Fermi surfaces of $RbTi_3Bi_5$. (e) Fermi surface in $k_z = 0.5$ plane. The black regular hexagon is the first Brillouin zone. The corresponding frequencies are given as the numbers in Tesla units.

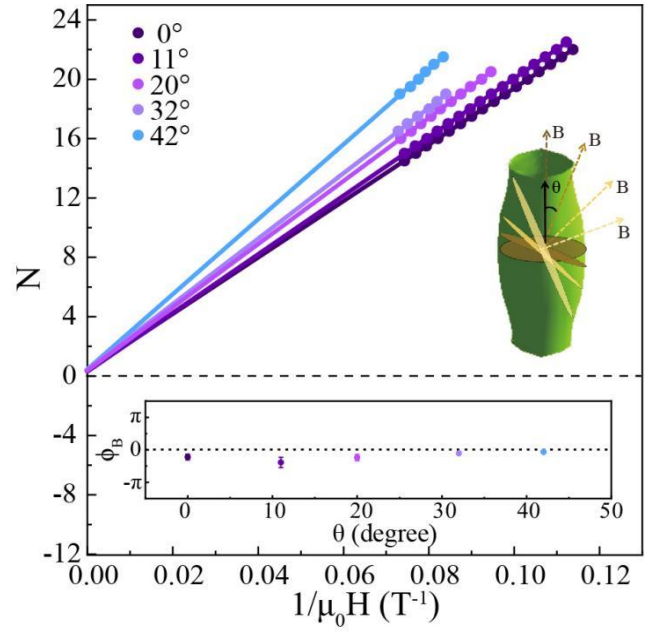

FIG. 4. Landau index N vs $1/\mu_0H$ for the frequency of $F_\beta$, derived from 2 K at various angles of the magnetic field deviating from the *c* axis. The inset is Berry phase $\phi_B$ versus the rotating magnetic field angle θ and the schematic is the Fermi pocket for different magnetic field directions